\documentclass[twocolumn,prb]{revtex4-1}

\usepackage{epsfig,amssymb,amsmath}

\usepackage[english]{babel}
\usepackage{epstopdf}

\begin{document}

\title{Dependence of the maximal superconducting current on resonance frequency in shunted Josephson junction}

\author{Yu. M. Shukrinov~$^{1,2}$}
\author{I. R. Rahmonov~$^{1,3}$}
\author{G. Filatrella~$^{4}$}

\address{
$^{1}$ BLTP, Joint Institute for Nuclear Research, Dubna, Moscow Region, 141980,  Russia\\
$^{2}$ Dubna State University, Dubna, Russian Federation\\
$^{3}$ Umarov Physical Technical Institute, TAS, Dushanbe, 734063 Tajikistan\\
$^{4}$ Department of Sciences and Technologies and Salerno Unit of CNISM, University of Sannio, Via Port'Arsa 11, I-82100 Benevento, Italy
}

\date{\today}

\begin{abstract}

We have investigated the phase dynamics and IV-characteristics of shunted Josephson junctions coupled to an $LC$ circuit. When the Josephson frequency $\omega_J$ is close to the eigen frequency $\omega_{rc}$ of the coupled resonance circuit, the IV-characteristic demonstrates an additional $rc$-branch. We have investigated  the features of the $rc$-branch and of the superconducting current component for different values of the resonance frequency. It has been found that the maximal value of the superconducting current depends on the resonance frequency and  that it increases when the end point of the $rc$-branch approaches the critical current.
The dependence of the maximal superconducting current on the resonance frequency at different values of the dissipation parameter is peculiar, for the optimized maximum appears to be independent of the system parameters within $1\%$.

\end{abstract}
\maketitle

\section{Introduction}
One of the simplest and effective methods to influence the features of Josephson junctions (JJ) is to shunt the device with  linear elements as capacitances, inductances, and resistances  \cite{likharev86}.
The JJ together with these elements form different resonant circuits, which are the essential constituents of modern superconducting electronics.
When the Josephson frequency is comparable with the frequency of the coupled resonant circuit, the Josephson oscillations can be tuned to the external circuit frequency, and the synchronization can also be revealed through the IV-characteristic as a step, a hump, or a dip \cite{tachiki11}, or negative differential resistance branch \cite{Pedersen73,Pedersen14}.
Recently  \cite{hriscu2013}, shunted JJ have been suggested for the creation of electric current standard and to close the metrology triangle ``voltage-resistance-current''.
Shunting the JJ by $LC$ elements can be also used to synchronize the Josephson oscillations in the BSCCO high temperature superconductors stacks and to increase the power of the coherent  electromagnetic radiation in the terahertz  region \cite{tachiki11,lin11}.

An essential point of interest of the shunting of Josephson junction by LCR-cirquit is that it allows the contemporary presence of two frequencies for the same set of parameters, or birhythmicity, which is encountered in some biochemical systems \cite{decroly,morita,sosnovtseva,enjieu}, nonlinear electronic circuits \cite{zakharova,yamapi,ghosh,yamapi12}, and extended distributed systems \cite{gasagrande}. The experimental observation of birhythmic systems was discussed in Refs. \cite{ventura,gonzalez}. In this context the superconducting circuit consisting of Josephson junctions (JJs) coupled to a cavity \cite{hadley,filatrella92,ozyuzer} , as in Fig. 1, represents a preeminent example of birhythmic system that is also interesting for different applications.
The coupling among the junctions is supposed to be provided by a resonant cavity \cite{gross,grib06,grib13}, thus when all the junctions are entrained it is essential to have a large current in the cavity such that the junctions can been  locked  through the current in the resonator.
The state with a large current coexists with a state at lower power; the two states are clearly characterized by two different frequencies.
This is the essential feature of birhythmicity, the coexistence of two attractors characterized by two different amplitudes and frequencies; depending on the initial conditions, the system can produce oscillations at two distinct periods.
Since the attractors are locally stable, the system would however stay at a single frequency, the one selected by the choice of the initial state. Thus the system exhibits a hysteretic behavior: the displayed frequency depends upon the initial conditions. In the presence of noise the system can switch from one attractor to the other under the influence of the random term \cite{yamapi14}.

The existence of resonance features in the IV-characteristics in various systems of JJs with a resonance circuit was reported in a number of experimental and theoretical works (see \cite{likharev86,almaas02}, and references therein). A peak in the intensity of coherent electromagnetic radiation from a
two-dimensional system of JJs based on Nb/Al/AlOx/Nb was detected in Ref.\cite{barbara} at the synchronization of oscillations in different JJs, which is caused by the resonance of Josephson oscillations with the oscillations in the resonance circuit. The considered system has a high potential for applications in quantum metrology, as well \cite{hriscu2013,shukrinov-epl15}.

The effect of external electromagnetic radiation on the shunted system when  creates new phenomena when the induced vertical voltage interval (usually called Shapiro Step, SS) appears in correspondence of the resonance circuit branch.
Particularly, when the corresponding SS is on the resonance branch, the amplitude dependence of its widths  is determined by the effective frequency which varies in a wide interval \cite{shukrinov-epl15}.

Much attention to the Josephson junctions is nowadays due to the appearance of the novel physical
phenomena and a variety of modern applications \cite{fu08,alicea12,rokhinson12,wilczek15,avriller15} . Devices based on the Josephson junctions are used to register Majorana particles, whose bound states are topologically protected and might be used for quantum computation. Shunting of Josephson junctions in such devices and external radiation gives an additional degree of freedom to manipulate and control their parameters and characteristics. Shunting of intrinsic Josephson junctions in high-temperature superconductors, that are intensively studied today, leads to the synchronization of oscillations of the superconducting currents in this system and increase the power of the tunable coherent electromagnetic radiation in the terahertz frequency region \cite{welp13}.

The role of  $LC$s shunts has been recently underlined in coupled systems \cite{Shukrinov12-JETP}, Shapiro steps \cite{Shukrinov15-EPL} and coupled stacks \cite{Shukrinov16-WEPJ}. The present paper focus on the behavior of the superconducting current component in  JJ shunted by an $LC$ element.
It has been found that the superconducting current in the state corresponding to the resonant branch depends on its frequency.
The value of the maximum current is determined by the proximity of the end point of the rc-branch to the critical current.
When dissipation is decreased, the maximum shifts towards the larger resonance  frequencies and its width  broadens.

The paper is organized in the following way. In Sec. 2, we introduce the model and the system of equations that is used in our numerical simulations.  In Sec. 3, we present the results of simulation, analysis of the IV  characteristics at different shunt parameters and demonstrate general features  of the $rc$-branch and behavior of superconducting current. The variation of the resonance frequency and $\beta$-dependence of the maximal superconducting current are discussed in Sec. 4.
We present a plot of the $rc$-branches at different value of the resonance frequency and show the resonance frequency dependence of the maximal superconducting current.
In Sec. 5, we analyze the endpoint of $rc$-branch at fixed $rc$-frequency  Finally, we discuss the obtained results and come to the conclusions.

\section{Model and Method}
\begin{figure}[htb]
\centering
\includegraphics[height=50mm]{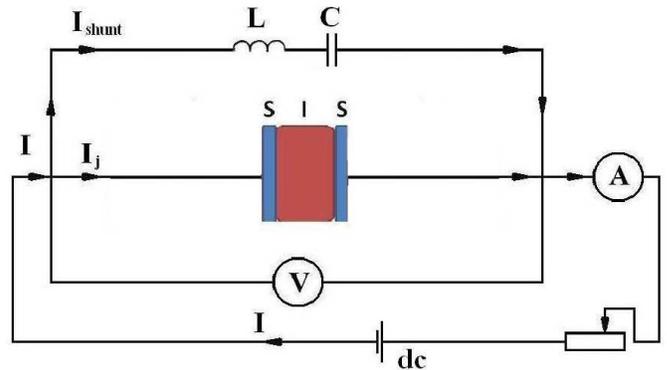}
\caption{ System of JJs shunted by inductance $L$, capacitance $C$ and resistance $R$. $S$ and $I$ denote the superconducting and dielectric layers, respectively. }
\label{1}
\end{figure}

To stress the main features of the shunted systems we concentrate on the description of the electric schema presented in Fig.~\ref{1}.

The total current through the system is a sum of the currents through JJ, $I_j$, and the current through the shunt elements, $I_{sh}$. To describe the properties of the considered system, we use the usual RCSJ model which captures its main features and very well reflects the experimental results \cite{likharev86}.
In the framework of  RCSJ-model \cite{likharev86}, the current $I_j$ is determined by
$$I_{j}=C_{j}\frac{dV}{dt}+I_{c}sin \varphi+\frac{\hbar}{2eR_{j}}\frac{\partial \varphi}{\partial t},$$
where $C_j$,  $R_j$, and $I_c$ are used for the capacitance, resistance and critical current of JJ,  while  $V$ and  $\varphi$ denote the voltage and the phase difference, respectively.
The current $I_{sh}$ is determined by  $I_{sh}=C {\partial u_{c}}/{\partial t}$, where  $C$ and $u_c$ are the shunt capacitance and voltage.
In the parallel shunt connection the voltage of the JJ is equal to the sum of voltages on the shunted elements, i.e.,
$$V = LC\frac{\partial^2 u_c}{\partial t^2}+RC\frac{\partial u_c}{\partial t}+u_{c}.$$
In this paper we do not take into account the shunt resistance to stress the main effects of the shunting resonator.

In normalized units the corresponding system of equations  can be written in the form \cite{shukrinov-jetpl12}
\begin{equation}
\label{system_eq3}
\left\{\begin{array}{ll}
\displaystyle\frac{\partial \varphi}{\partial t}=V
\vspace{0.3 cm}\\
\displaystyle \frac{\partial V}{\partial t}=I-\sin \varphi-\beta\frac{\partial \varphi}{\partial t}-C\frac{\partial u_{c}}{\partial t}
\vspace{0.3 cm}\\
\displaystyle \frac{\partial^{2} u_{c}}{\partial t^{2}}=\frac{1}{LC}\bigg(V-u_{c}\bigg).
\end{array}\right.
\end{equation}
Here the bias current $I$ is normalized to the critical current $I_c$ of JJ, time $t$- to the inverse plasma frequency $\displaystyle\omega_{p}=\left( {{2eI_{c}}/{C_{j}\hbar}} \right)^{1/2}$, the voltages $V$ and $u_c$ (voltage at the capacitance $C$) are normalized to $\displaystyle V_{0}={\hbar \omega_{p}}/{2e}$;  the shunt capacitance $C$ - to the capacitance of the JJ $C_j$, and shunt inductance $L$  to $(C_{j}\omega_{p}^{2})^{-1}$.
In the system of equations (\ref{system_eq3}) we introduce a dissipation parameter $\displaystyle \beta= \left( {1}/ {R_{j}}\right) \left( {\hbar}{2eI_{c}C_{j}}\right)^{1/2}= \left( {{\beta_{c}}}\right)^{-1/2}$, $\beta_{c}$ being  the McCumber parameter.
Here we present the results for the underdamped JJ with $\beta = 0.5 \div 0.2$.
We note that JJ together with the LC-elements form a parallel resonance circuit with its eigenfrequency \cite{likharev86}
\begin{eqnarray}
\omega_{rc}=\sqrt{\frac{1+C}{LC}}
\label{w}
\end{eqnarray}
and a series resonance circuit with
 \begin{eqnarray}
\omega^s_{rc}=\sqrt{\frac{1}{LC}}
\label{w}
\end{eqnarray}
In this paper, we concentrate mostly on the parallel resonance. The details of the model and simulation procedure are presented in Refs.\cite{likharev86,Shukrinov16-WEPJ}.

\section{The rc-branch and superconducting current}
\begin{figure}[htb]
\centering
\includegraphics[height=65mm]{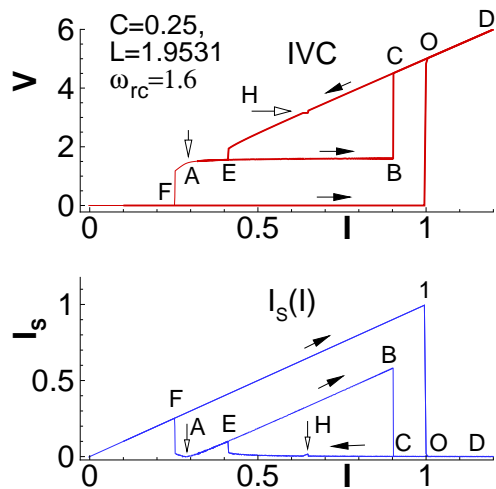}
\caption{(a) Demonstration of the  $rc$-branch $AB$ in IV-characteristics of shunted JJ with $L=1.9531$, $C=0.25$ with bias current sweep along $01ODEAEBCDEF0$; (b) The I-dependence of the superconducting current $I_s=\sin{\varphi}$, obtained at the same sweep.}
\label{2}
\end{figure}

As we have mentioned in the Introduction, the specific feature of IV-characteristic of shunted JJ is a resonance circuit branch.  Figure~\ref{2} shows how to change the bias to reach the branches we are interested in. We do so to closely reproduce the experimental procedure, that cannot select initial conditions.
We thus start from $I=0$, where the initial condition with all variables set to $0$ is guaranteed, and slowly change the parameters, keeping the last value of the dynamic variables as the initial values for the next value of the bias current $I$. In this way, we explicitly give the prescriptions to reach the resonant branch and the resistive states. Figure~\ref{2}(a) demonstrates the IV-characteristic of a  JJ shunted with $L=1.9531$, $C=0.25$, that is a resonant circuit of frequency $\omega_{rc}=1.6$.
The calculations are based on the system of equations  (\ref{system_eq3}) as the bias current sweeps along the path  $01ODEAEBCDEF0$. At variance with non-shunted  IV-characteristics that exhibits hysteresis at $\beta<1$ (see for example  [\onlinecite{buckel04}]), the IV-characteristic of shunted JJ demonstrates  $rc$-branch due to the resonance of Josephson oscillations with the eigenmode of the resonance circuit,  $\omega_{J}=\omega_{rc}$.
In the case presented in Fig.\ref{2}(a), the $rc$-branch, beginning from the hollow arrow is  $AB$  (see below).

The position of the  $rc$-branch endpoint corresponds to the voltage $V=1.6$, as expected for the eigenfrequency at $N=1$, $L=1.9531$, $C=0.25$ in Eq. (\ref{w}).
This proves that  the $rc$-branch  is due to the aforementioned resonance.
Changing the $LC$ parameters, the $rc$-branch move accordingly towards higher or lower frequencies. \cite{shukrinov-jetpl12}.

The dependence of the superconducting current $I_s$ as a function of the  bias current $I$ along the sweep path of Fig.\ref{2}(a) is shown in Fig. \ref{2}(b), where we use the same letters to mark the same bias points on the IV-characteristics and  supercurrent contribution $I_s(I)$. In the zero voltage state $I_s$ increases from $0$ to $1$, and then sharply decreases of $3$ or $4$ orders of magnitude along the resistive branch until it vanishes altogether.
When the current is decreased from point $D$,  the supercurrent $I_s$ slowly increases again, demonstrating a small hump in $rc$-harmonic region around $V=3.2$ (shown in IV-characteristic by  $H$), and then essentially increases at transition to $rc$-branch (point $F$).
Proceeding further,  $I_s$ decreases up to transition to the state with $V=0$ (point $F$).
Point A where $I_s$ is minimal is at the  beginning of the $rc$-branch.
Along the $rc$-branch,  $I_s$ increases almost linearly and exhibits a maximum in point $B$, where $I_s=0.52$.

\begin{figure}[htb]
\centering
\includegraphics[height=65mm]{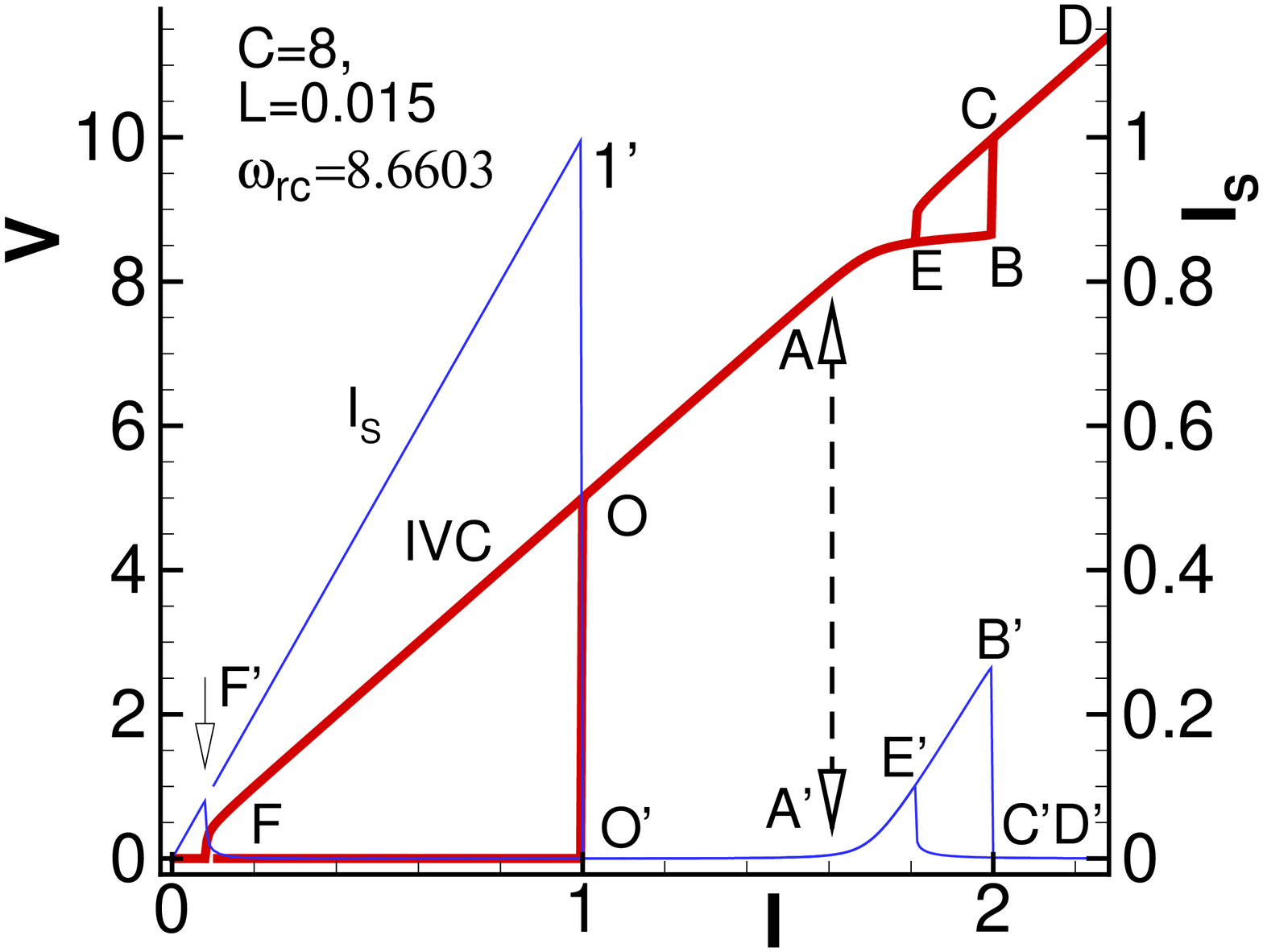}
\caption{IV-characteristic (thick line) of shunted JJ at $C=8$, $L=0.015$ together with $I_s(I)$ dependence (thin line), obtained by sweeping along $01OBCD  EF0$.}
\label{3}
\end{figure}

The close behavior is observed at another parameters of the resonance circuit.   In Fig. \ref{3} we show IV-characteristic and $I_s(I)$ dependence at $L=0.015$, $C=8$ ( $\omega_{rc}=8.6603$) by sweeping along $01OBCD EF0$. Figure~\ref{3} as well demonstrates the correspondence of the eigen frequency to the branch endpoint position at $V=8.6603$. Thus, the Fig. \ref{3} demonstrates that the behavior observed in Fig. \ref{2} is generic.

At variance with the previous case, the $rc$-branch is outside the hysteresis region.
Another important difference is worth underlying: the maximal superconducting current is essentially smaller and consists of $I_s=0.27$ (point $B'$), instead with $I_s=0.52$ at $\omega_{rc}=1.6$.
A question appears: which parameters determine the maximum value  $I_s$ can reach along the $rc$-branch? We will return back to this question in next sections.

As we have stressed above, in this work we focus on the superconducting current
\begin{equation}
I_s = \langle\sin{\varphi}\rangle,
\label{Is}
\end{equation}
the average current flowing through the nonlinear Josephson element. Such current contributes to the total bias current together with the current through the resistor (the capacitor current being obviously zero on average).
The contribution of the current through the Josephson element is essential to lock the junction either to an external drive, as elucidated by the the standard Bessel function approach \cite{Braiman80,Kautz81} or a power balance method \cite{Filatrella93}.
In fact, in both approaches the extension of the Shapiro steps is given by the dc current that flows through the Josephson element. This current (power, in the power balance approach) can be either positive or negative (depending on an arbitrary phase), and the larger the dc contribution of the Josephson current the larger the height of the Shapiro step.

\begin{figure}[htb]
\centering
(a)\\
\includegraphics[scale=0.4]{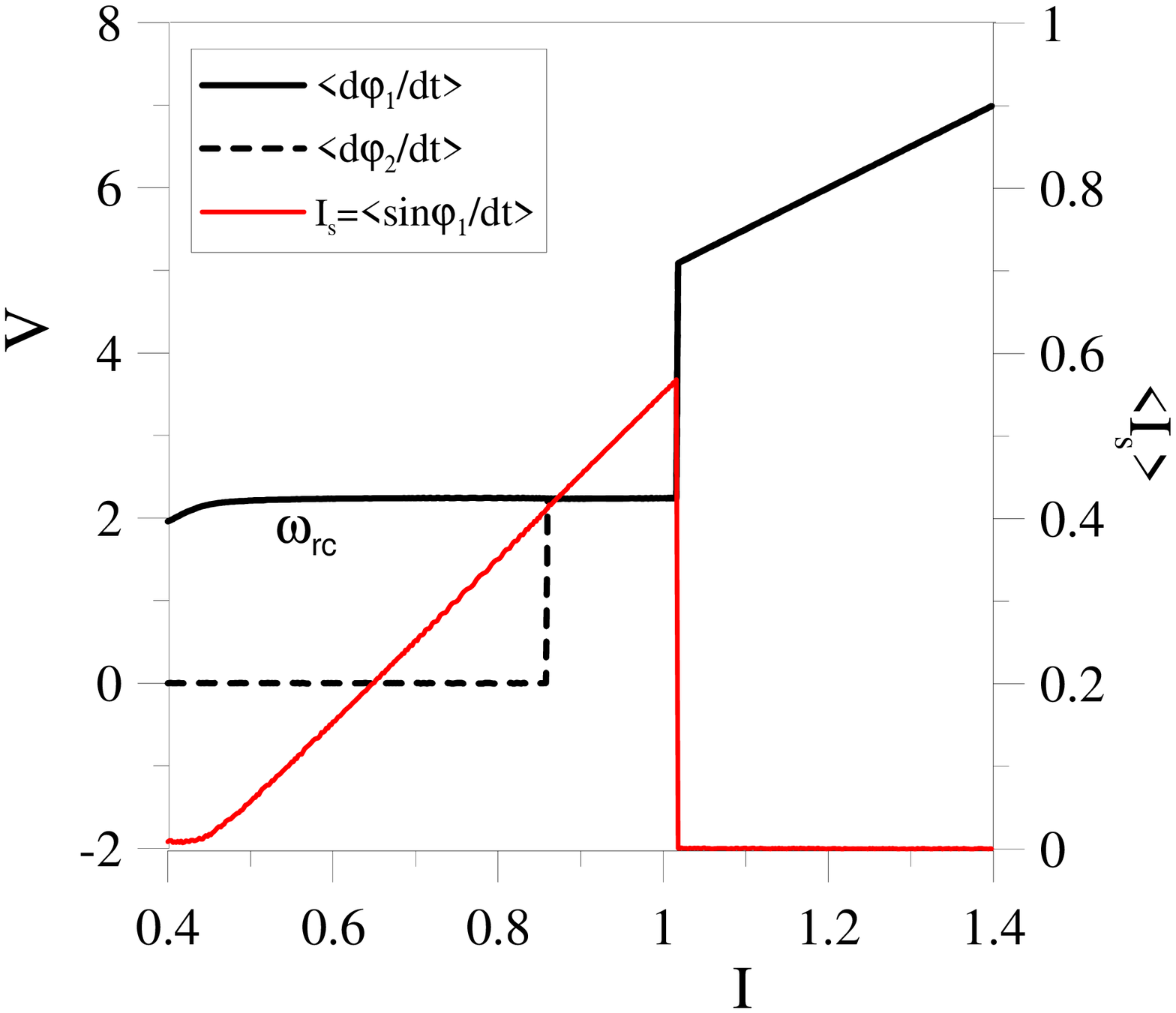}\\
(b)\\
\includegraphics[scale=0.4]{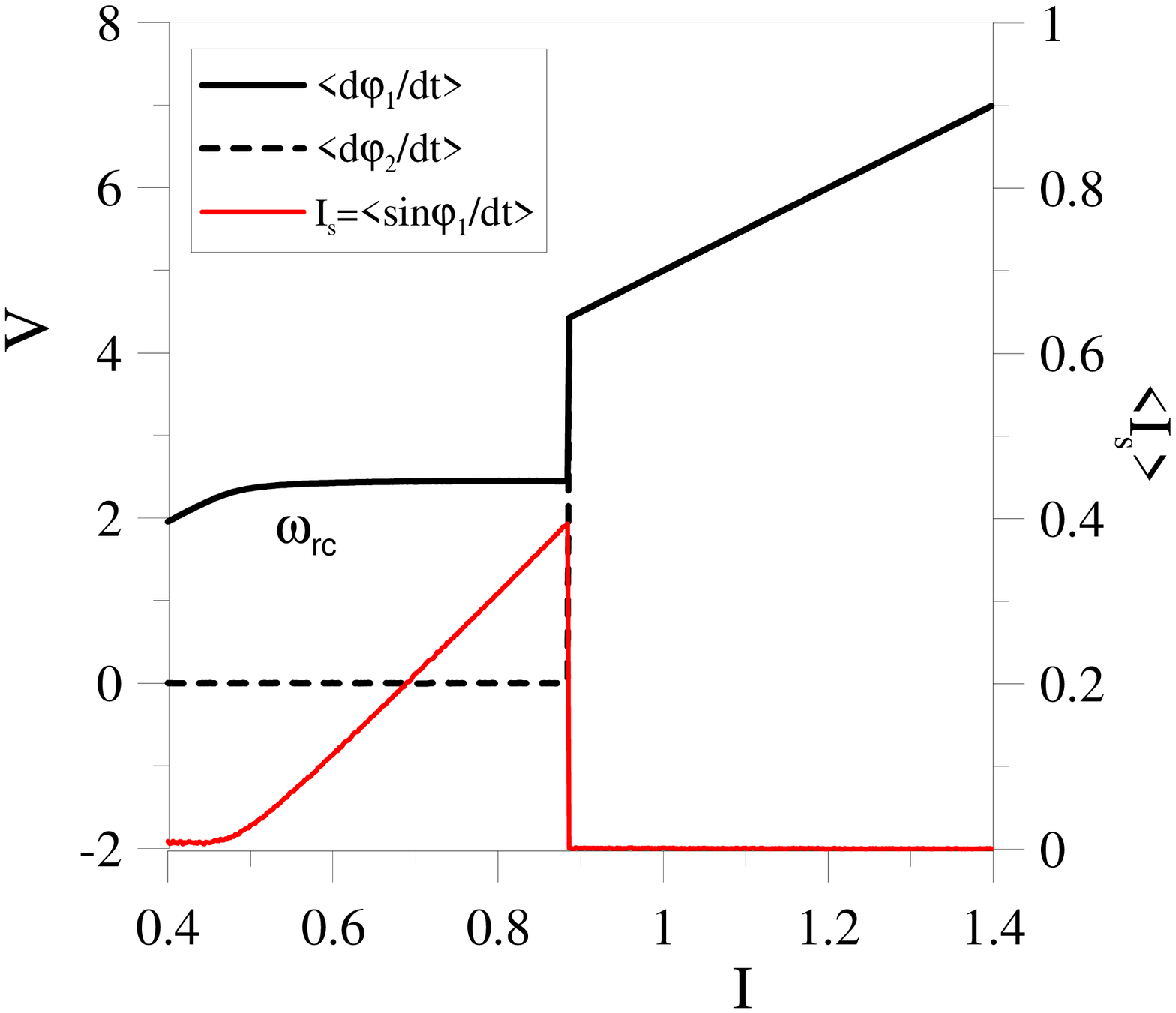}
\caption{The IV characteristics of two coupled and shunted JJ , as per Eq.(\ref{system_eq_series}), with increasing bias current sweep  on the branch $0D$ of Fig. \ref{2}.
The black solid and dashed lines denote the average voltages of the two junctions, respecitvely.
The red line denotes the syperconducting current, as per Eq.(\ref{Is}).
The locking, that is marked by the region where the two voltages coincide, only occurs when the supercurrent passes $0.4$.
In (a)  $L=1.9531$, while in (b) $L=1$ .
In both figures  $C=0.25$, $\beta = 0.2$, $\delta = 0.1$.
The $I$-dependence of the superconducting current $I_s =  \langle \sin{\varphi} \rangle $ is reported on the right vertical axis.
}
\label{4}
\end{figure}

The dc current through the Josephson element is also a key feature of JJ arrays synchronization. In fact it is this term that activates the oscillations responsible of the coupling among the junctions, that would otherwise be just uncoupled \cite{Hadley88}.
This is shown in Fig. \ref{4} where we report the simulations for two JJs  with slightly different critical currents: junction $1$ critical current is $10\%$ higher that the critical current of junction 2. Thus the normalized critical current of one junction reads $1+\delta$, while the other is as usual normalized to $1$.
The two junctions are in series, so that both are biased by the same current $I_j$ in Fig. \ref{1}, and described by the following set of equations:
\begin{equation}
\label{system_eq_series}
\left\{\begin{array}{ll}
\displaystyle\frac{\partial \varphi_1}{\partial t}=V_1
\vspace{0.3 cm}\\
\displaystyle\frac{\partial \varphi_2}{\partial t}=V_2
\vspace{0.3 cm}\\
\displaystyle \frac{\partial V_1}{\partial t}=I-\sin \varphi_1-\beta\frac{\partial \varphi}{\partial t}-C\frac{\partial u_{c}}{\partial t}
\vspace{0.3 cm}\\
\displaystyle \frac{\partial V_2}{\partial t}=I-(1+\delta) \sin \varphi_2-\beta\frac{\partial \varphi_2}{\partial t}-C\frac{\partial u_{c}}{\partial t}
\vspace{0.3 cm}\\
\displaystyle \frac{\partial^{2} u_{c}}{\partial t^{2}}=\frac{1}{LC}\bigg(V_1+V_2 - u_{c}\bigg).
\end{array}\right.
\end{equation}

In Fig. \ref{4}(a) we display the simulations with two values of the normalized inductance $L$.
It is evident that the junctions are synchronized (that for the JJ relation implies the same voltage) only at high values of the term $I_s = \langle \sin{\varphi}\rangle$.
 In fact synchronization, the overlap of the voltages, only occurs for $I_s  0.4$.
Such value of the supercurrent is only reached for the larger value of $L$ in (a), and it is not reached for the lower value in (b).
It is therefore important to optimize the system to obtain the largest values of this superconducting current $I_s$,  that is the subject of next sections.

\section{Resonance frequency and $\beta$-dependence of the maximal superconducting current}

At variance with the previous case, the $rc$-branch is outside the hysteresis region.
Another important difference is worth underlying: the maximal superconducting current is essentially smaller and consists of $I_s=0.27$ (point $B'$), instead with $I_s=0.52$ at $\omega_{rc}=1.6$.
A question appears: which parameters determine the maximum value  $I_s$ can reach along the $rc$-branch?

\begin{figure}[htb]
 \centering
\includegraphics[height=60mm]{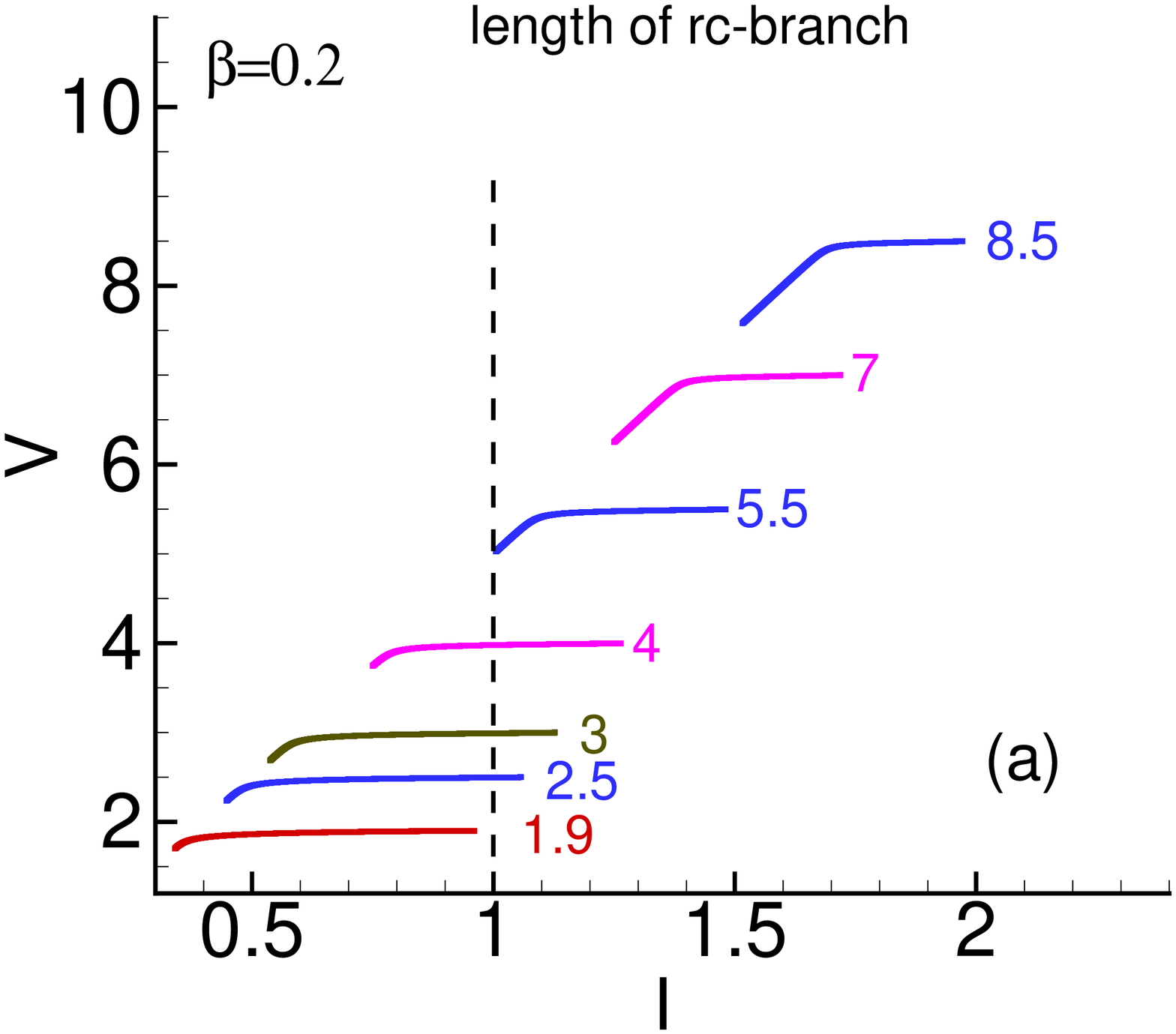}
\includegraphics[height=60mm]{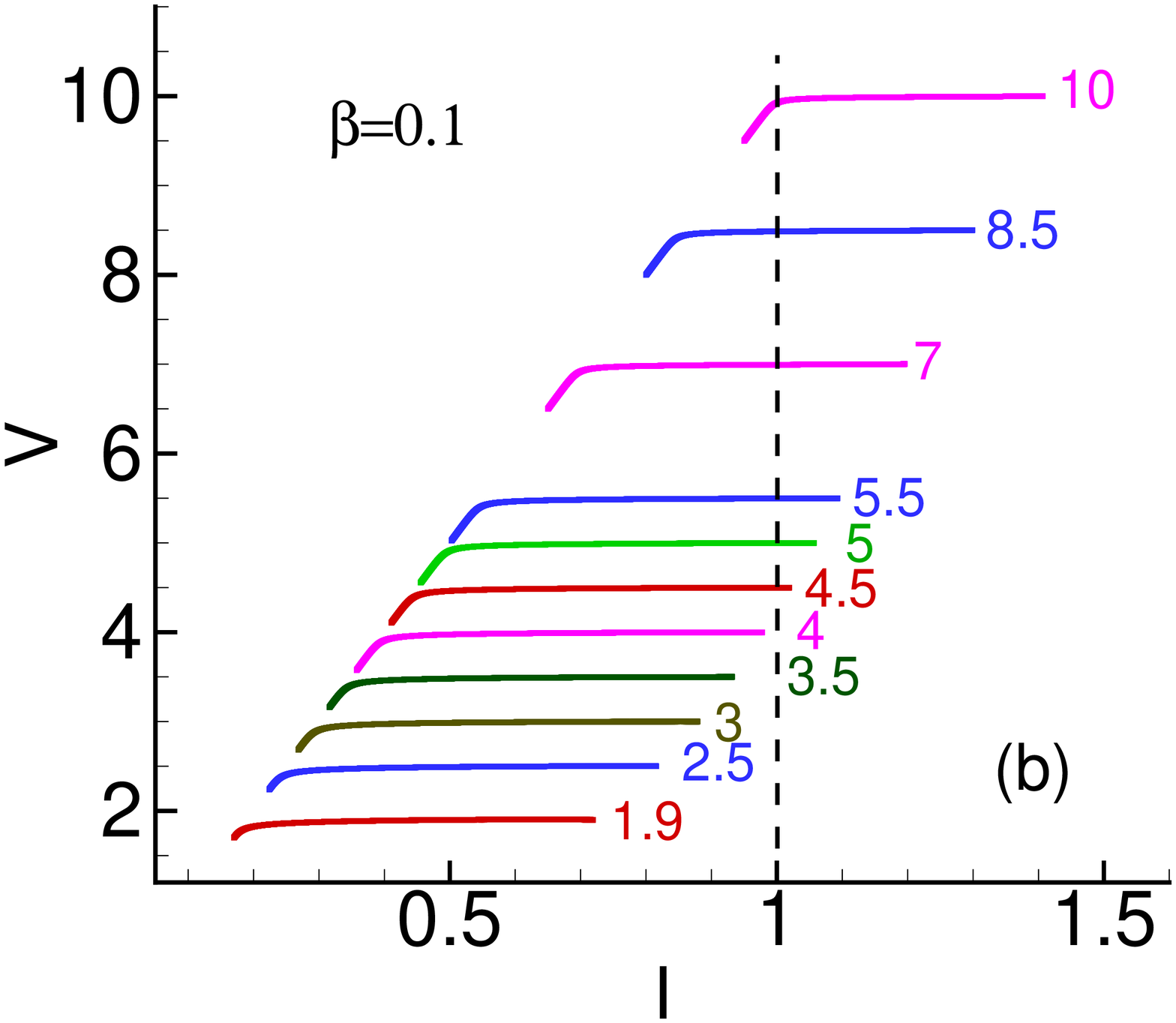}
\includegraphics[height=60mm]{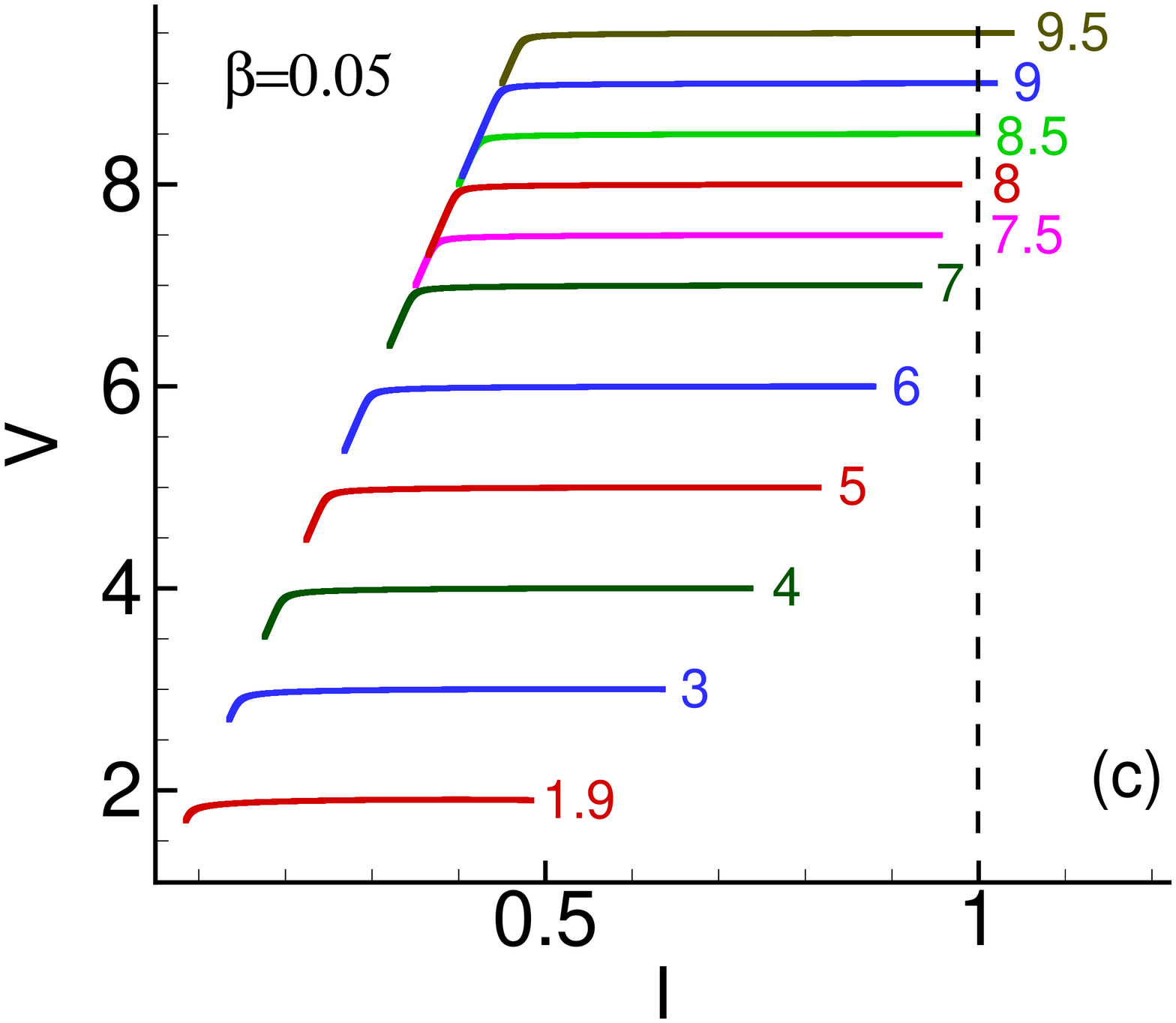}
\caption{$rc$-branch at different value of the resonance frequency: (a) $\beta=0.2$; (b) $\beta=0.1$; (b) $\beta=0.05$. Dashed line determines the value of bias current $I=1$. }  \label{5}
\end{figure}

To find the answer, we have made detailed investigations at different resonance frequency and dissipation parameter $\beta$.
The results, collected in Fig. \ref{5}, refer to the resonance branch (($A-B$ in Fig. \ref{2}).
Figure  \ref{5} shows the dc-current $I$, that contains the nonlinear term $I_s$, see Eq.(\ref{Is}).

The $rc$-branches at $\beta=0.2$ and the resonance frequency in the interval $(1.9, 8.5)$  are presented in Fig.\ref{5}(a).
We underline that the horizontal part of $rc$-branch changes non-monotonically with resonance frequency: the increase is  followed by a decrease, as it is seen very clearly in Fig. \ref{5}.
The results for $\beta=0.1$ and $\beta=0.05$ are presented in Fig. ~\ref{5}(b,c).
In all cases the $rc$-branch with maximal horizontal part is realized.
We note that the variation of the $rc$-branch is obtained changing the shunt inductance and keeping the capacitance fixed, $C=0.25$.

\begin{figure}[htb]
 \centering
\includegraphics[height=50mm]{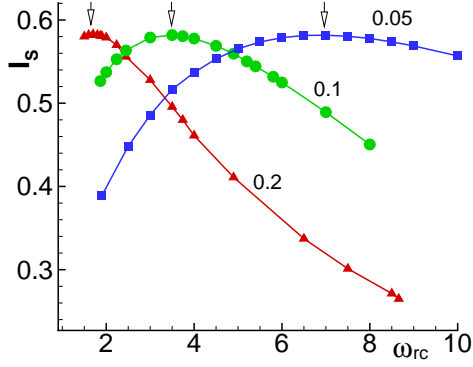}
\caption{Resonance frequency dependence of the maximal superconducting current. Squares ($\beta=0.05$), circles ($\beta=0.1$), and triangles ($\beta=0.2$) show the value of the resonance frequency.}
 \label{6}
\end{figure}

In  Fig.~\ref{6} it is reported the resonance frequency dependence of the maximal superconducting current in $rc$-branch at different dissipation  $\beta=0.05, 0.1, 0.2$.
Each point in this Figure corresponds to the maximum supercurrent along the branches reported in Fig. \ref{5}. We have systematically observed that this maximum is reached at the largest dc bias current $I$. The nonmonotonical dependence of the $I_s(\omega_{rc})$ is related  to the fact that the $I_s$ -part in the total current is increasing with approaching $I_c$, but sharply decreased when $I_c$ is passed.

All figures demonstrate the maximum of $I_s$ in the interval between $I_s=0.581$ and  $I_s=0.583$. In case $\beta=0.2$  it is  $I_s=0.5828$ and at $\beta=0.1$ and $\beta=0.05$ it is  $I_s=0.5819$ and $I_s=0.5817$, respectively.  Decrease of dissipation (increase of hysteresis region in IV-characteristic) leads to the shift of maximum for larger $\omega_{rc}$.

The maximal value of $I_s$ on $rc$-branch is determined by how close its end point $I_{end}$ to the critical current $I_c$. In considered cases the maximal values of  $I_s$ are realized in the following way: at  $I_{end}=0.9227$ in case  $\beta=0.2$, and at  $I_{end}=0.9318$ in both cases $\beta=0.1$ and  $\beta=0.05$.

\section{Endpoint at fixed $rc$-frequency}

\begin{figure}[htb]
 \centering
\includegraphics[height=50mm]{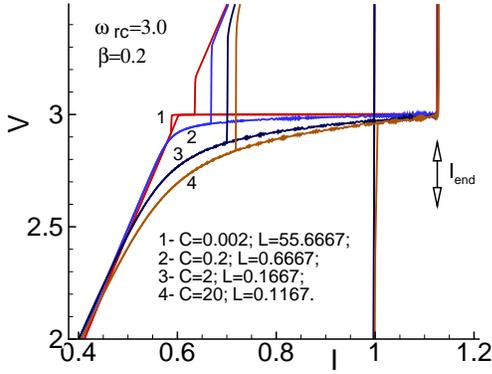}
\caption{Independence of the $rc$-branch end point on shunt parameters at fixed resonance frequency  $\omega_{rc}=3.0$.}  \label{7}
\end{figure}

We have also addressed another question: the determinant of the the size of the $rc$-branch.
To this extent   it has we have changed the shunt inductance and capacitance in a wide interval, leaving the constant of resonance frequency, $\omega_{rc}=3.0$.
The results are shown in Fig. ~\ref{7}, where  it is evident that at fixed resonance frequency the end point of the $rc$-branch does not change.
However, the values of $L$  and $C$ influence on the onset of $rc$-branch.
At small $C$, the $rc$-branch  is a practically a straight  line, that  bends when the capacitance $C$  increases and its onset moves towards the region of small currents.

The reactance $X$ of this system is expressed by
\begin{equation}
\frac{1}{X}=\frac{1}{X_{C}+X_{L}}+\frac{1}{X_{C_{j}}}
\label{res1}
\end{equation}

with $\displaystyle X_{C}=\frac{1}{i\omega C}$, $ \displaystyle X_{L}=i\omega L$ and $\displaystyle X_{C_{j}}=\frac{1}{i\omega C_{j}}$, so

$$\displaystyle X=i\omega \frac{1-\omega^2LC}{C+C_{j}-\omega^2LCC_j}$$

In this system can be observed two type of resonances: serial and parallel.

The serial resonance in the system is realized when reactance is equal zero;  when this is the case,  Eq. (3) gives the resonance frequency.
The formula leads to the maximum of ac current through this circuit branch and to the minimum of the ac voltage.
The case when the admittance is equal zero, corresponds to the parallel resonance with  frequency (4) that leads to the maximum ac voltage.

\begin{figure}[htb]
 \centering
\includegraphics[height=35mm]{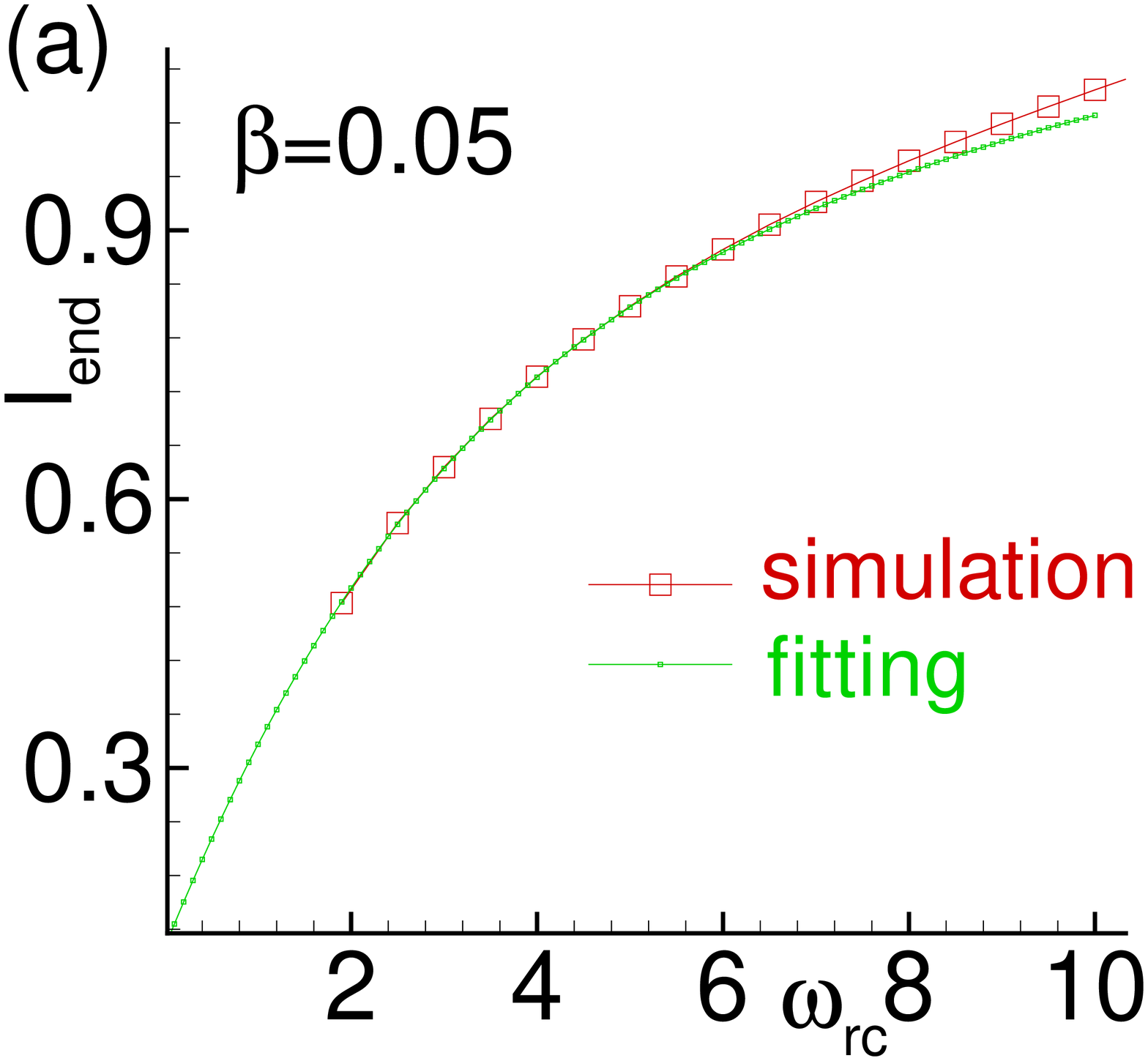}
\includegraphics[height=35mm]{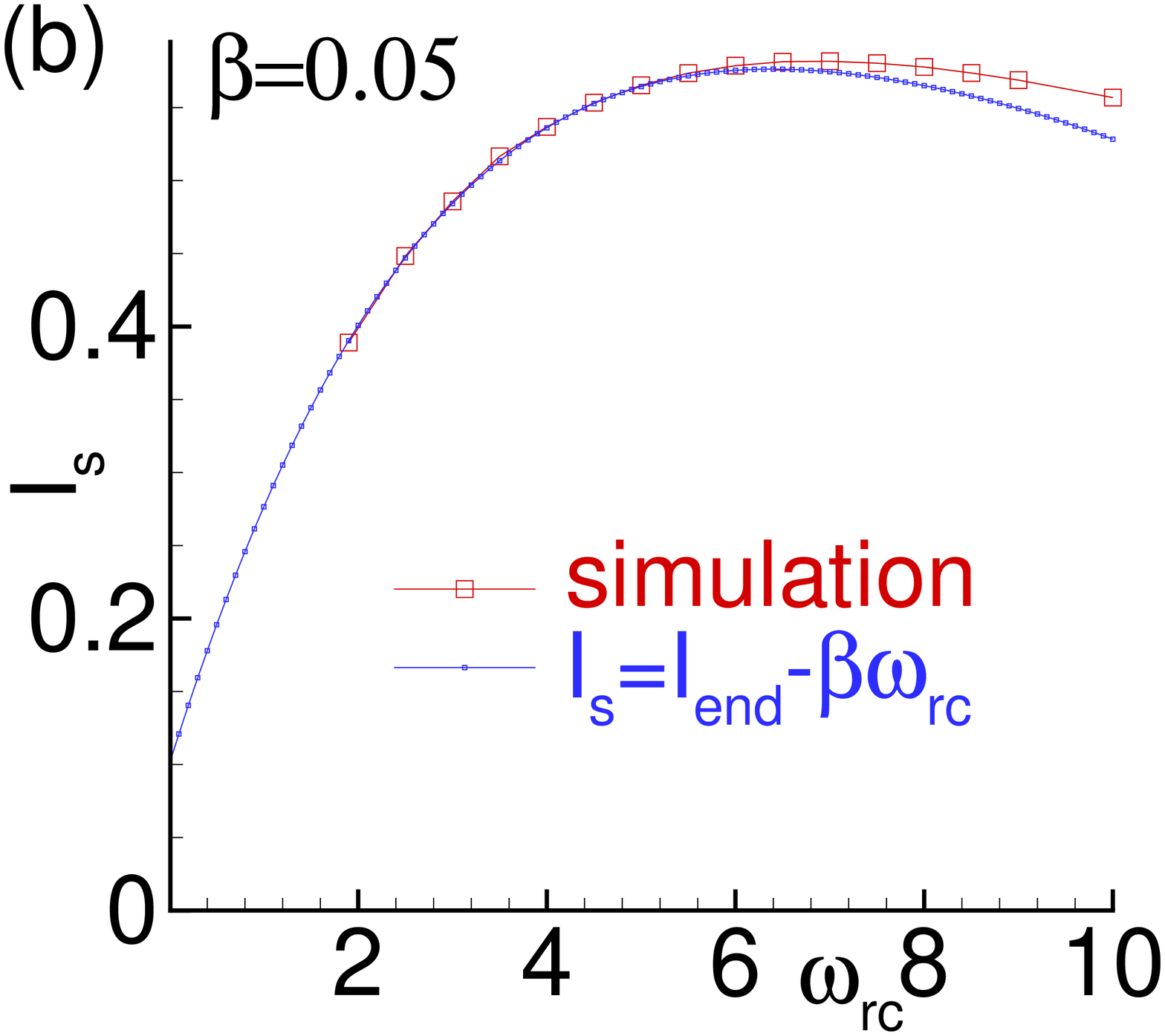}
\includegraphics[height=35mm]{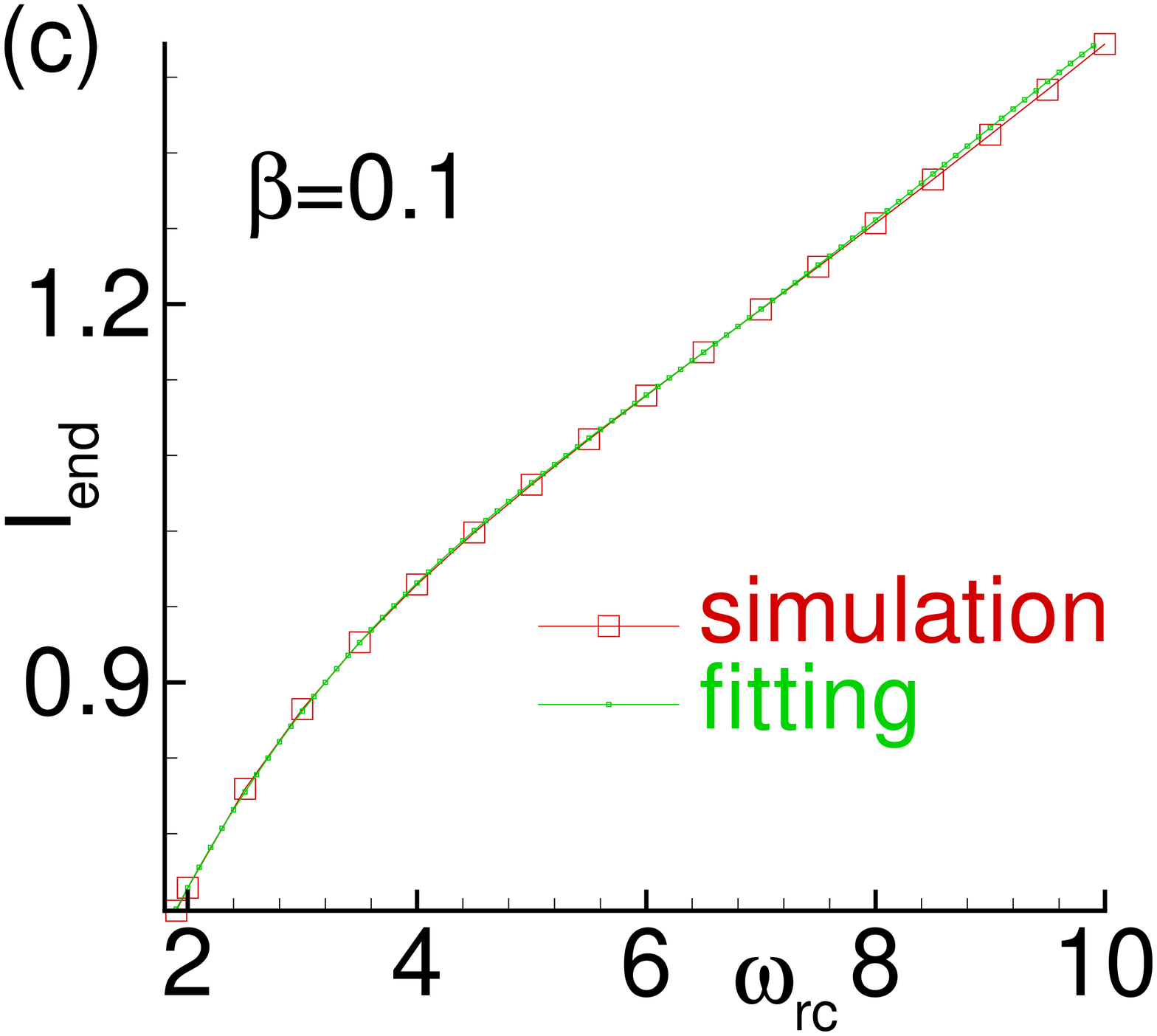}
\includegraphics[height=35mm]{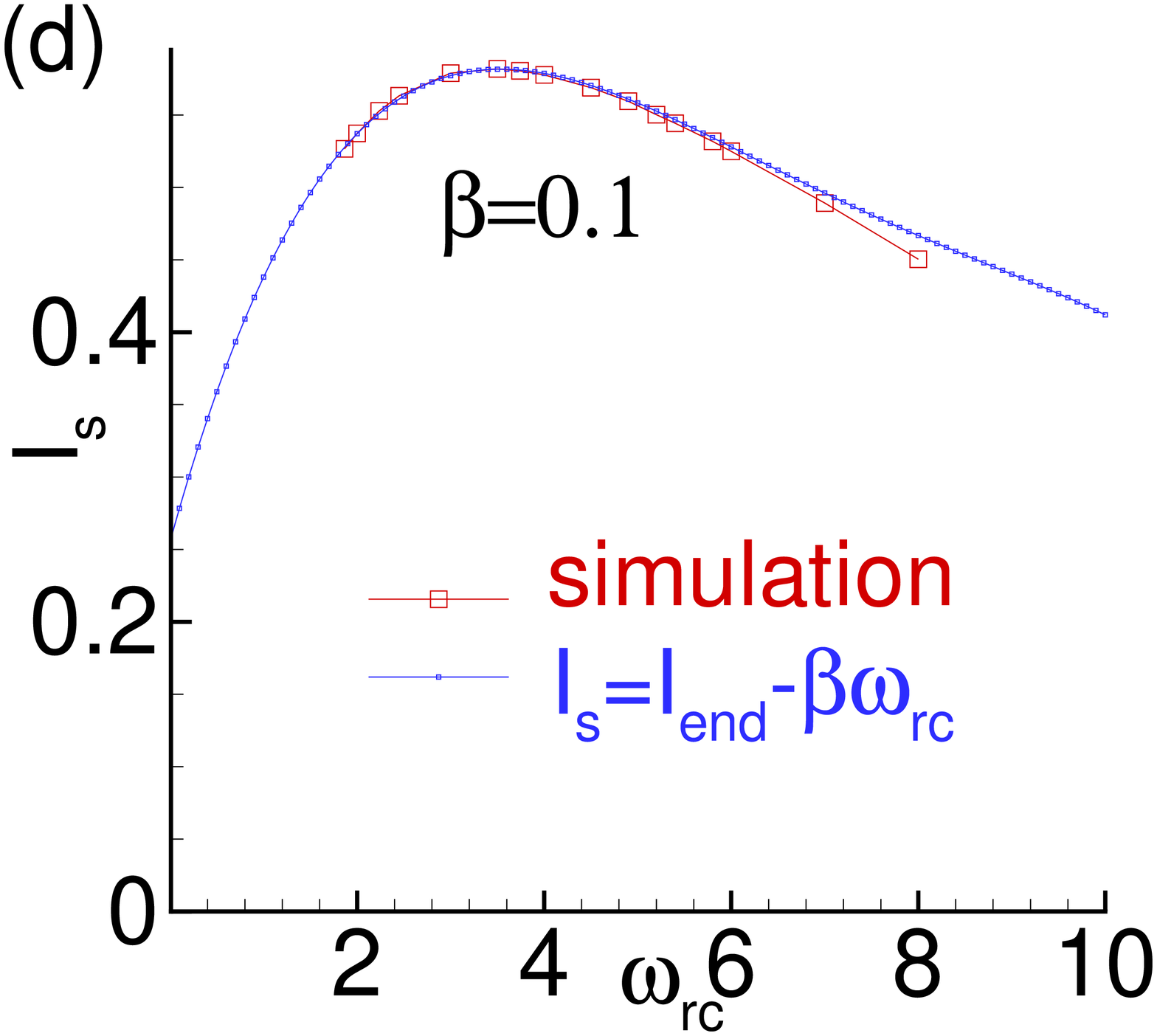}
\includegraphics[height=35mm]{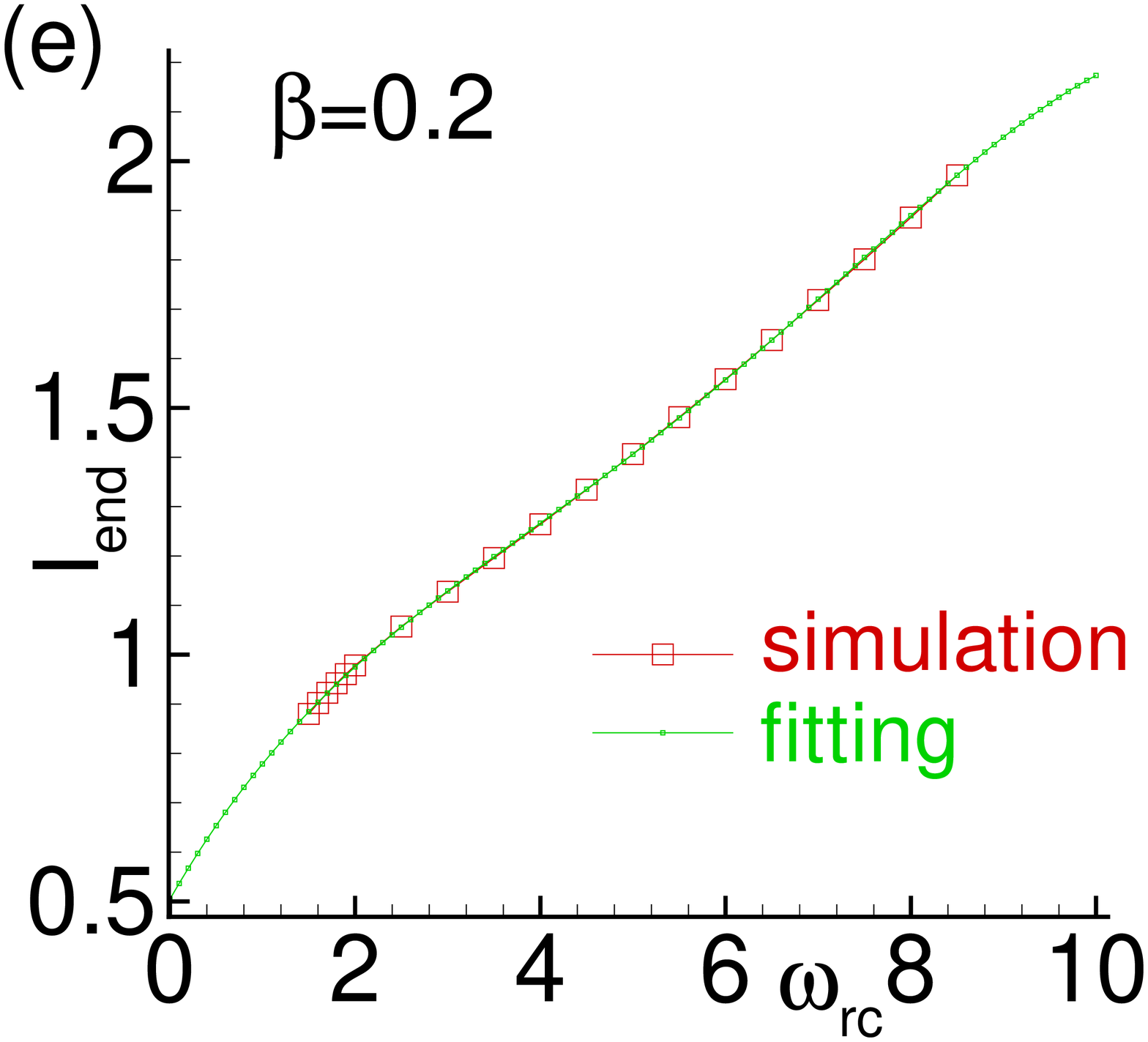}
\includegraphics[height=35mm]{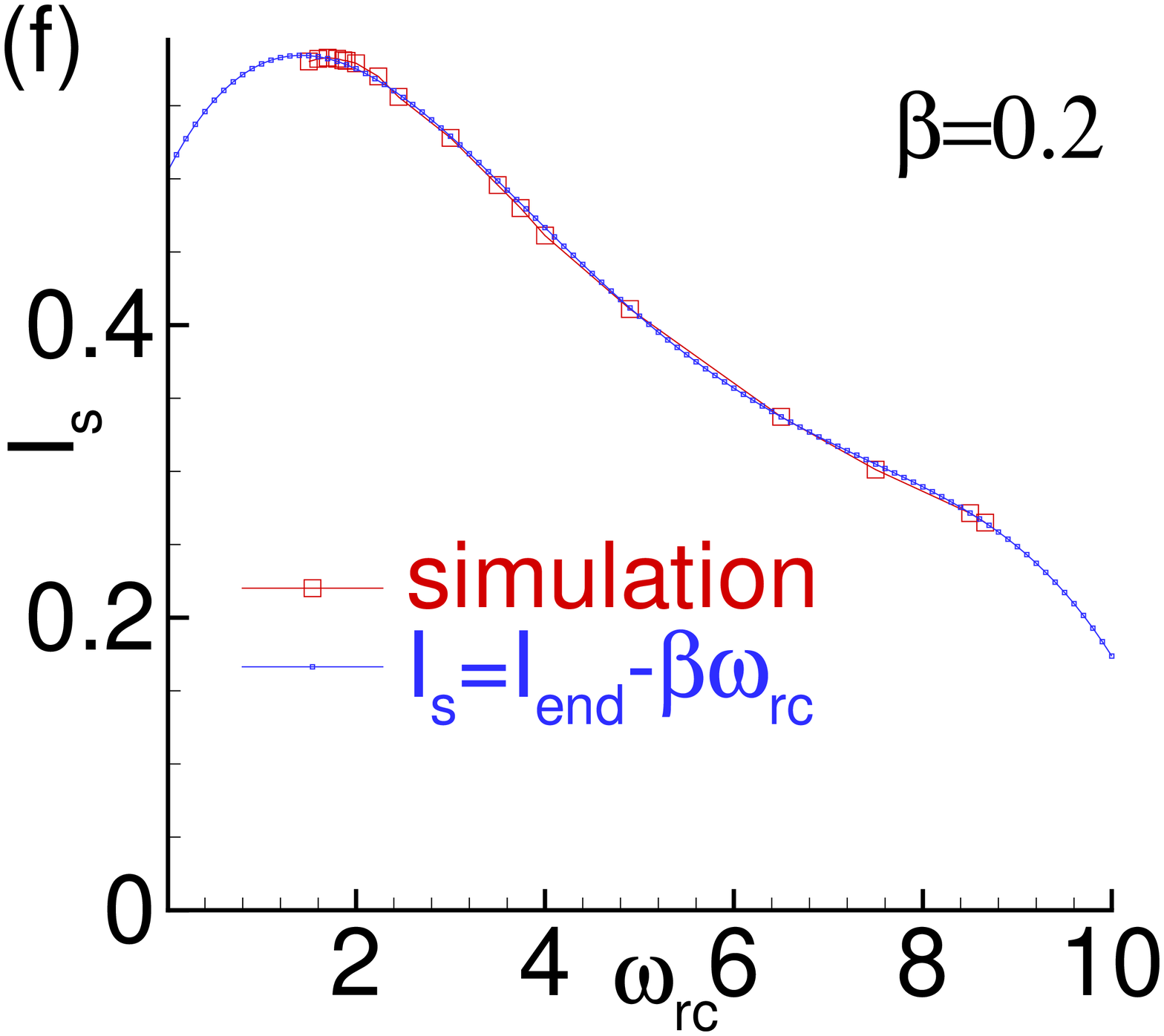}
\caption{Dependence of $I_{end}$ and  $I_{s}$ on $\omega_{rc}$ at different dissipation parameters together with results of fitting by formula (\ref{polinom}).}
\label{8}
\end{figure}

With a parallel circuit resonator the equivalent resistance ($L_{sh}$, $C_{sh}$, $C_j$ in series) diverges, and so does the bias current (that is a sum of the superconducting and normal currents).
At the resonance (using our normalization) we may write $I_s+\beta \omega_{rc}=I_{end}$, where we take into account that $I_n=\beta \omega_{rc}$.
So, at fixed $I_{end}$ the maximum of the superconducting current is determined by value of $I_n$ at the resonance.
The approximation $I_s+\beta \omega_{rc} \simeq I_{end}$ is in very good agreement with numerical simulations.
The equation $I_s+\beta \omega_{rc}=I_{end}$ has two unknown functions of bias current (or $\omega_{rc}$). To clarify it, we plot the simulated dependence of $I_{end}$ and  $I_{s}$ on $\omega_{rc}$ at different dissipation parameters in Fig.\ref{8} together with results of fitting by formula (\ref{polinom}).

\begin{table}
\caption{Fitting coefficients}
\label{tabular:timesandtenses}
\begin{center}
\begin{tabular}{cccc}
Coefficients & $\beta=0.05$ & $\beta=0.1$ &$\beta=0.2$ \\
$A_{0}$      & 0.10061      & 0.25578     & 0.50439\\
$A_{1}$      & 0.25557      & 0.33344     & 0.32673\\
$A_{2}$      & -0.03179     & -0.05588    & -0.06008\\
$A_{3}$      & 0.00214      & 0.00511     & 0.00791\\
$A_{4}$      & -0.00006     &-0.00017     & -0.00035\\
\end{tabular}
\end{center}
\label{table}
\end{table}

For the parallel resonance the reactive impedance of the system tends to infinity.
Therefore, at the top of the resonance peak the bias current $I=I_{end}$ is a sum of $I_{s}$ and $I_{qp}$, i.e., $$I_{end}=I_{s}+\beta V$$. Then
\begin{equation}
I_{s}=I_{end}-\beta \omega_{rc}
\label{Is}
\end{equation}

From the simulations we have obtained that $I_{end}$  depends on $\omega_{rc}$.
Fitting of $I_{end}$ by a fourth order polynomial
\begin{equation}
I_{end}= A_{0} + A_{1}\omega_{rc} +A_{2}\omega_{rc}^{2}+A_{3}\omega_{rc}^{3}-A_{4}\omega_{rc}^{4}
\label{polinom}
\end{equation}
with coefficients presented in Table \ref{table} for different $\beta$ values is shown in Fig . \ref{8}.
Using Eqs. (\ref{Is}) and (\ref{polinom}) we find $I_{s}$.
The increase of $I_{end}$ with $\omega_{rc}$ and the decrease of $I_s$ at large $\omega_{rc}$ leads to the appearance of the maximum in the dependence $I_s (\omega_{rc})$.

\section{Summary and Conclusions}

We have investigated the phase dynamics and the IV-characteristics of $LC$ shunted Josephson junctions. Along the resonant step, we have found that the superconducting current contribution is a key feature for the locking of JJ with slightly different parameters. We have therefore investigated the features of  the $rc$-branch and of the corresponding superconducting current at different values of the resonance frequency. We have found that the maximal value of the superconducting current depends on the resonance frequency and it is determined by proximity of the $rc$-branch to the end point of the critical current. The dependence of the maximal superconducting current on the resonance frequency at different values of the dissipation parameter has been found to be almost the same for a wide range of the system parameters. A design fitting formula has been found to be effective for the range of practical JJ parameters.

The authors thank S. Pagano, K. Kulikov, V. Kornev, M. Kupriyanov for useful discussions. Yu. M. S. wishes to thank Prof. S. Pagano for kind hospitality in during his visit Italy, where part of this work was written. The reported study was funded by RFBR according to the research projects 15-51-61011 Egypt, 15-29-01217 and  16-52-45011 India.

\end{document}